# LineCAPTCHA Mobile: A User Friendly Replacement for Unfriendly Reverse Turing Tests for Mobile Devices

C.B Bulumulla
Department of Statistics and Computer Science
University of Peradeniya
Peradeniya, Sri Lanka
chamindabulumulla@gmail.com

R.G Ragel
Department of Computer Engineering
University of Peradeniya
Peradeniya, Sri Lanka
roshanr@pdn.ac.lk

*Abstract*— **As smart phones and tablets are becoming ubiquitous and taking over as the primary choice for accessing the Internet worldwide, ensuring a secure gateway to the servers serving such devices become essential. CAPTCHAs play an important role in identifying human users in internet to prevent unauthorized bot attacks. Even though there are numerous CAPTCHA alternatives available today, there are certain drawbacks attached with each alternative, making them harder to find a general solution for the necessity of a CAPTCHA mechanism. With the advancing technology and expertise in areas such as AI, cryptography and image processing, it has come to a stage where the chase between making and breaking CAPTCHAs are even now. This has led the humans with a hard time deciphering the CAPTCHA mechanisms. In this paper, we adapt a novel CAPTCHA mechanism named as LineCAPTCHA to mobile devices. LineCAPTCHA is a new reverse Turing test based on drawing on top of Bezier curves within noisy backgrounds. The major objective of this paper is to report the implementation and evaluation of LineCAPTCHA on a mobile platform. At the same time we impose certain security standards and security aspects for establishing LineCAPTCHAs which are obtained through extensive measures. Independency from factors such as the fluency in English language, age and easily understandable nature of it inclines the usability of LineCAPTCHA. We believe that such independency will favour the main target of LineCAPTCHA, user friendliness and usability.**

## I. Introduction

CAPTCHA is an endeavour to prevent bots from having unauthorized access to various types of computing services or hacking into confidential and sensitive information. This is achieved by generating and grading Artificial Intelligence (AI) problems which humans can pass easily while computer programs cannot or considerably unlikely to pass. The word CAPTCHA is an acronym known as Completely Automated Public Turing test to tell Computers and Humans Apart [1]. Preventing spam, protecting automated website registration from bots, prohibiting online poll hijacking and preventing dictionary attacks are some of the major applications behind the need for a CAPTCHA mechanism [17].

The main problem with most of the modern CAPTCHAs is their user friendliness. Most of the unbroken CAPTCHAs are hard to understand by humans or they take a long time to solve, thus making CAPTCHA passing a frustrating task.

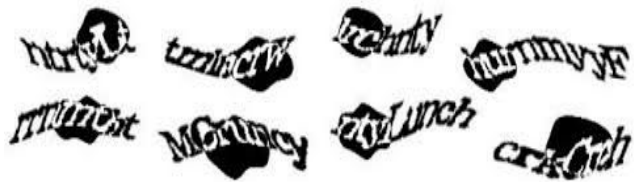

Fig. 1. Hard to solve text-based CAPTCHAs

Fig. 1. delineates such text-based CAPTCHAs which are the examples from the most frequently used CAPTCHA methods. CAPTCHAs such as reCAPTCHA [2, 19] uses techniques like scanned text that the Optical Character Recognition (OCR) technology has failed to interpret, and therefore results in bringing forward CAPTCHAs which are complicated. When it comes to mobile devices, they are usually even harder to see and interpret, because of the displays of limited size and limited means of user interaction.

In this paper, we propose LineCAPTCHA as an alternative CAPTCHA method specifically for non-user-friendly CAPTCHAs specifically targeting mobile devices. Although, the idea of LineCAPTCHA was proposed earlier [18], it was neither implemented on a mobile device nor systematically modelled and evaluated. Therefore, in this paper we are, for the first time, proposing LineCAPTCHA as a mobile friendly CAPTCHA system and also formulating a method of evaluation for such a system. We claim the two as our contributions in this paper.

## II. Background

The first notion of CAPTCHA, even though it was not named CAPTCHA then, was brought up by Moni Naor [7]. Since then CAPTCHA has attracted researchers from AI, cryptography, signal processing, and computer vision. Luis von Ahn et al. [1] of Carnegie Mellon University came up with the term CAPTCHA, after wards actively researching on CAPTCHA while developing numerous CAPTCHA systems such as Gimpy [23], Pix [24], reCAPTCHA [2] and AnimalPix [26].

All CAPTCHAs can be classified into text-based, image-based, audio and video based categories. Text-based CAPTCHAs represent individual characters, which are randomly tilted, shifted up and down and placed so as to overlap one another in order to obfuscate them. Image-based CAPTCHAs are often about drawing, labelling, rotating or identifying images. They contain systematic distortions such as dithering, noise and quantizing. Audio-based CAPTCHAs are specifically focused on users with visual impairment.

These CAPTCHAs measure the ability of recognizing spoken language with random noise and distractions. Video-based CAPTCHAs filter users as they identify certain fragments of the played video.

### III. LITERATURE SURVEY

ReCAPTCHA [2, 28], one of the most widely used text-based CAPTCHAs, provides CAPTCHAs to quite a number of high profile sites such as Facebook, Ticketmaster and Craigslist. Earlier, when the distortions done were low, the users were able to easily identify the characters, which lead to high breakability of reCAPTCHA. Therefore, the present reCAPTCHAs are made harder hence the users are unable to solve them easily. Fig. 2. depicts two such hard to solve reCAPTCHAs.

Confident CAPTCHA [4] is a popular image-based CAPTCHA used in mobile devices. The major drawback is that this is not resistant to no-effort attacks. No-effort attack is a set of random guesses that would work out well without taking any effort to solve the AI problem. Although the allowed breakability for random inputs in a strong CAPTCHA is less than 0.02%, Confident CAPTCHA has more than 10% breakability. Additionally, there can be situations where the user does not understand the meaning of the word suggested to click [Refer Fig.3.a].

NuCAPTCHA [5], which is presented in Fig. 3.b, is one of the popular video CAPTCHAs. Although, a NuCAPTCHA is only 300 x 250 pixels large, it uses about 750KB data. Battery is a scarce resource on mobile devices and NuCAPTCHA drains batteries of mobile devices by forcing them to play video CAPTCHA movies. Elie Burstein, a researcher at Google has broken the NuCAPTCHA video scheme [6] by extracting frames of the video and then analysing and finally identifying the objects.

In 2008, a CAPTCHA mechanism for mobile devices was proposed, which is appropriate for use on a touch-screen device, named Drawing CAPTCHA [8]. In Drawing CAPTCHA, the user must connect the three red diamonds with lines, to form a triangle. Fig.4. gives an overview of an algorithm in order to break Drawing CAPTCHAs [16].

In 2010, VidoopCAPTCHA [9], an image-based CAPTCHA was proposed that allows users to look for specific images and then asks them to type in the letters from those images. This CAPTCHA method provided easy to use protection until two researchers Michele Merler and Jacquilene Jacob was able to break it [3]. They used images downloaded from Flicker for training data and 200 VidoopCAPTCHA images as test data. With image processing techniques such as feature extraction, image splitting and character recognition, they were able to break 9 VidoopCAPTCHAs per hour.

In 2012, an animated CAPTCHA, Vappic, consists of six alphanumeric characters stored inside a patterned cylinder that rotates in the centre was proposed. One of the major drawbacks was that users who could not touch type would look down from the screen to find the keys and would lose their place in the rotating animation. Therefore they would have to wait for the animation to rotate around again to solve it [10].

Asirra [11] is a Microsoft research project that asks users to distinguish between cats and dogs out of 12 photographs. From the user studies it can be seen that this CAPTCHA is solved by humans 99.6% of the time within 30 seconds. In mobile devices, since this is compounded on small screens users get confused because the small image made it difficult to identify details.

In 2013, TapCHA [12] presents a novel CAPTCHA design which features a hybrid challenge combining text recognition based reading comprehension task and shape and motion puzzles. Shape puzzle suffers from no-effort attacks.

We consider that drawing lines on a mobile screen be easily performed by human and therefore propose LineCAPTCHA, where curves and drawing lines are used to tell computers and humans apart. In LineCAPTCHA the user is asked to find a randomly generated curve and to draw on top of it. In this paper we have discussed how the implementation of the LineCAPTCHA is done, security aspects of LineCAPTCHA and how its role as an alternative for the existing CAPTCHA mechanisms targeting mobile devices. The users can submit the CAPTCHA for evaluation if they are satisfied with what they have drawn or they have the opportunity to cancel the drawn curve and get a fresh LineCAPTCHA reverse Turing test.

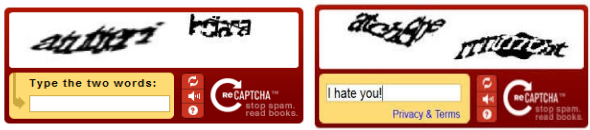

Fig.2. Hard to solve ReCAPTCHAs [28]

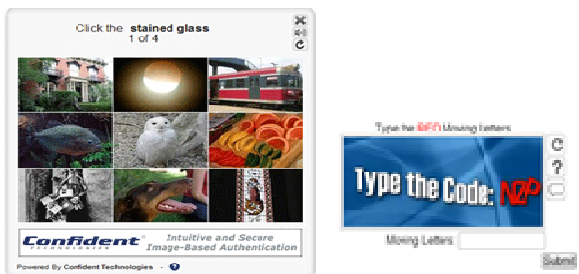

(a)                  (b)

Fig.3. Confident CAPTCHA and NuCAPTCHA

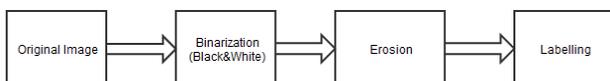

Fig.4. Overview of an algorithm to break drawing CAPTCHA [16]

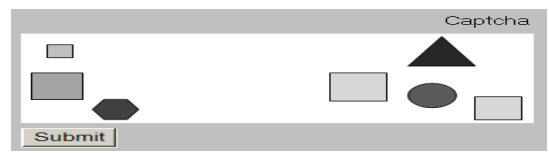

Fig.5. Sample image of TapCHA

## IV. METHODOLOGY

LineCAPTCHA for mobile is developed under android development environment. We will elucidate the methodology of LineCAPTCHA under 3 main phases and they are background generation, cubic Bezier curve generation and evaluation. First the background of the CAPTCHA is generated with randomly selected 8 images from the database. Next, control points of the cubic Bezier curve are randomly generated and according to the points the curve will be created. After that the user has to identify the Bezier curve and draw on top of it and submit. Data sets for the generated and drawn points will be stored and sent to another android activity for evaluation. Finally the result will be displayed on the screen determining whether the user has passed or failed the CAPTCHA.

### A. Background Generation

A noisy background is developed using three different schemes. The main objective is to create a background with appropriate level of distractions to provide a good human interactive proof challenge. Firstly, images are created using segmented English alphabetical letters and numbers from 0 to 9. Care was taken to match the size of the curve segments which will be drawn by the user and the size of character segments which will be in the images. Twelve such images were stored in the image database and the size of a single image is around 7KB.

Secondly, a 2-dimensional Grid View was setup in order to structure the background to hold the distracted images. We developed the grid view to hold 8 images (the number of images in the background can be varied if need arise) with zero padding between adjacent grids, which is brought out as a single final image.

Finally, a method was developed to randomly select unique set of images from the database and fill the grid view.

### B. Cubic-Bezier Curve Generation

Cubic-Bézier curves [13] were chosen, as such higher degree curves are more computationally expensive to evaluate so that these will enhance the security aspects of LineCAPTCHA. Cubic-Bézier curves flow from a start point to a destination point, with their curvature influenced by the 2 transitional control points [14].

Also it provides more security as some of the control points do not fall on the curve and therefore it is hard to regain the line without having at least two control points. Length of the curve can be changed according to security requirements by altering the arbitrary number, while keeping the concerns on the user's easiness to draw. Alternatively, higher order curves such as Spline polynomials, Cubic curves can also be used, which comprises of diverse geometric shapes and parameters.

$$B(t) = (1-t)^3 P_0 + 3(1-t)^2 t P_1 + 3(1-t) t^2 P_2 + t^3 P_3, t \in [0,1] \quad (1)$$

Equation (1) shows the Cubic-Bezier curve. We randomly generated values for the four control points ($P_0$, $P_1$, $P_2$ and $P_3$). These points will define the shape of the curve. Shapes vary to a wide range from loops, arcs, S-curves [Ref. Fig. 10]. The curve starts from $P_0$ in the direction of $P_1$, and arriving from the direction of $P_2$, stops at $P_3$. Two separate equations are used to derive values for x and y points on the curve. Arbitrary number (t) defines the length of the curve.

### C. Statistical Evaluation

After the curve is generated, the coordinate points of the curve are stored in an array list. We call it the curve data set. While the user is drawing on top of the curve, the coordinate points of the user drawing are also stored into an array list using a separate android thread. We call it the drawn data set. Evaluation processes in a separate Android activity.

Prior to a thorough statistical evaluation, there is some simple mathematical verification done in order to verify whether the user has drawn correctly on the curve. We propose this to make the evaluation system computationally efficient, so that careless attempts of users are easily rejected by a simple mathematical calculation without going through the statistical evaluation.

The main concept behind the evaluation is statistical hypothesis testing [21]. The following two hypotheses were defined considering the population mean as a parameter.

**Null Hypothesis**: Population means of the datasets are equal
**Alternative Hypothesis**: Population means are not equal

Level of Significance, the criterion of judgment upon which the decision is made regarding the condition stated in the null hypothesis, is taken at the 99% confidence level (Refer Fig. 8.). This interval is chosen to increase the user friendliness of the CAPTCHA. It should be noted that this confidence level can be changed to adjust the hardness of LineCAPTCHA according to the security measures needed at the deployed computer system. For instance, a 90% level of confidence can be used in a higher security system where the test value should be between ±1.645 in order pass through [22].

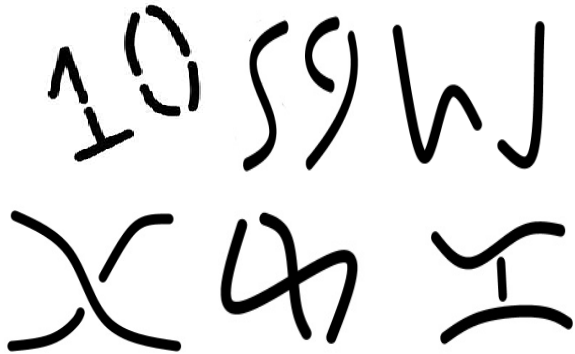

Fig.6. Images used in the background generation

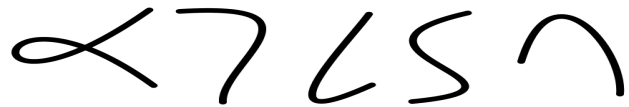

Fig.7. Higher order Bezier curves have lots of vacations which are also comfortable with user.

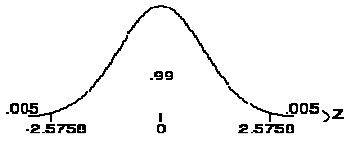

Fig. 8. 99% Confidence interval

The test was conducted using two sampled z test [15]. In order to use this test, the data under test should be normally distributed. We were unable to find an efficient normality testing statistical application for android to be integrated with our application and therefore implemented a separate program for doing so. This version is based on an algorithm originated from Shapiro-Wilk's normality test [25, 27]. Results of the normality program were further verified with the statistical software R console and we could conclude that the program we developed accurately tests for normality. We noted that, even without testing for normality, the system verified the users comparatively well. Sample data sets we are using have more than 30 elements and moreover a transformation for the data set can make the data set converge more towards normality.

$$z = \frac{(\bar{x}_1 - \bar{x}_2) - d_0}{\sqrt{\frac{\sigma_1^2}{n_1} + \frac{\sigma_2^2}{n_2}}} \quad (2)$$

Equation (2) shows how to calculate the z test statistic. $\bar{x}$ refers to the mean of the two sample data sets. In our case, $d_0$ is equal to 0 since we are testing for equal means. $\sigma$ is the standard deviation and n is the sample size of the two data sets. If the z values for the sample data sets are within ± 2.5758 [Refer Fig. 8.] then the LineCAPTCHA is considered as passed, otherwise failed.

## IV. RESULTS AND DISCUSSION

### A. User Experiment of LineCAPTCHA

We have conducted a user survey consisting of 40 candidates, differing from factors such as age, computer literacy, and literacy. The aim of the survey was to provide LineCAPTCHA a real user exposure and to evaluate the developed CAPTCHA system.

During the user experiments, two versions of LineCAPTCHA were presented to the users and they are short-LineCAPTCHA [Refer Fig. 9.], and Long-LineCAPTCHA [Refer Fig. 10.]. In long-LineCAPTCHA a single full Bezier curve is generated whereas in short-LineCAPTCHA only parts of the full Bezier curve are generated. Bezier curve is segmented to 3 different portions making the user identify 3 smaller curve segments instead of the long curve. Background images are created separately for the two LineCAPTCHA versions since Bezier curve that is generated in both versions differs from each other. In short-LineCAPTCHA, the background images contain small character segments which are similar to Bezier curve segments in characteristics like size, colour and shape. In long-LineCAPTCHA, the images consist of much longer character segments to match the single long Bezier curve generated.

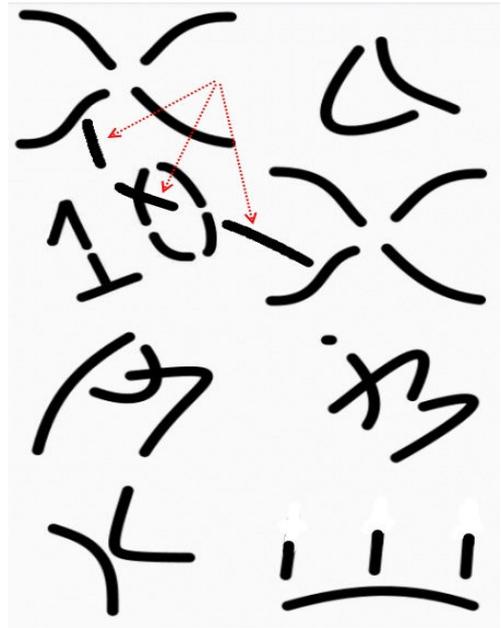

Fig.9. Example short-LineCAPTCHA

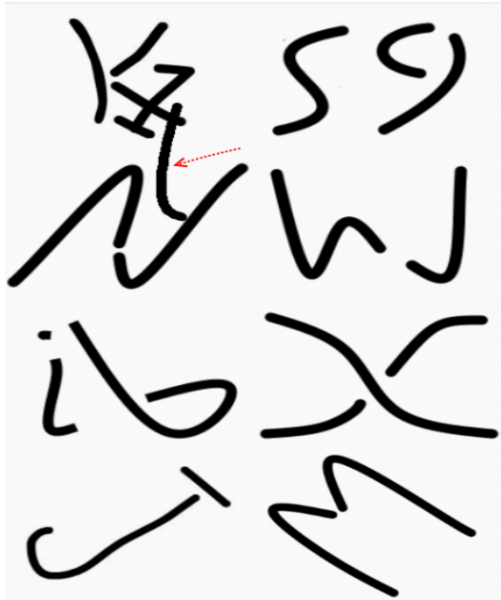

Fig. 10. Example long-LineCAPTCHA

These two versions were deployed on a Samsung Galaxy S Duos S7562 with a 1GHz Cortex-A5 processor. This android device has 480 x 800 pixels, 4.0 inches (~233 ppi pixel density) [20] display which is an average sized mobile display. In order to have unbiased results for small and large display devices this average sized device was chosen.

From the survey, all the users gave a good feedback regarding the general idea behind LineCAPTCHA, which is drawing on top a simple curve. They considered it as highly user friendly rather than deciphering stressful characters and numbers which might end up getting their eyes strained. Fig.

11. presents the evaluation results between the user friendliness of the two versions, and it is clear from the chart that long-LineCAPTCHA is more user friendly than short-LineCAPTCHA. About 85% of the candidates preferred LineCAPTCHA over ReCAPTCHA, while the remaining 15% consisted of users who were not aware about reCAPTCHA and users who had very low computer literacy. We could say the usability of this system is high because even those who have never solved a CAPTCHA before were able to solve LineCAPTCHAs regardless of factors like age, nationality and computer literacy. Most of the users tend to make mistakes at the two ends of the curve. Candidates were able to pass through LineCAPTCHA in less than 10 seconds after they get familiar with the LineCAPTCHA environment.

With the user-experience gained from the survey, we would like to impose certain standards to LineCAPTCHA to maintain the degree of quality in the system. The user should see what he/she is drawing since they tend to make fewer mistakes and focus more on drawing when they see what they are drawing. User should be allowed to redraw whenever needed, with a fresh background and a curve.

### B. Security Aspects of LineCAPTCHA

User friendliness of LineCAPTCHA can be processed according to the security measurements needed for the deployed system. In other words, LineCAPTCHA is capable of providing high security (with confusing background images, short-LineCAPTCHA where 3 curve segments are to be found rather than one, a higher significance level for evaluation) standards for a secure website to protect from dictionary attacks, and a low security LineCAPTCHA (with less amount of distraction integrated with only a long-LineCAPTCHA with a lower significance level) can be implemented to prevent spammers commenting on blog spots. Here we elaborate some of the security aspects of LineCAPTCHA.

There should be certain prominent features of LineCAPTCHA which will reduce the chances of breaking LineCAPTCHA from basic image processing techniques and they are discussed below.

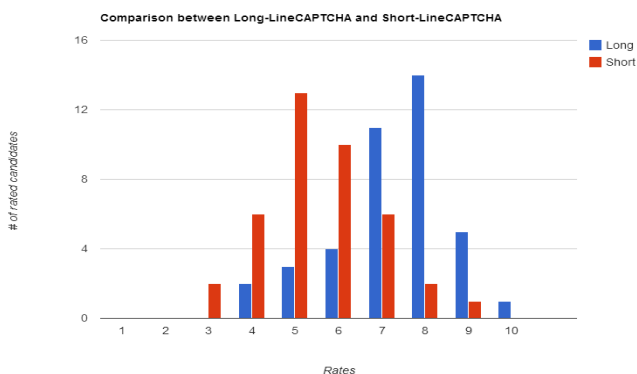
Fig.11. Evaluation of long- and short-LineCAPTCHAs

- **Resistant to colour filtering and thresholding**: filtering images by colour components can be done using different techniques in spaces such as RGB colour space and HSI space, which will simplify most complicated images to be easily broken. We use black and white images to prevent attacks with colour filtering.

- **Protection against segmentation analysis**: Background objects and the Bezier curve should not be easily segmented by techniques such as watershed segmentation, K-means clustering or texture filters.

- **Erosion and dilation**: LineCAPTCHA resists attacks from erosion and dilation since it contains very similar characteristics among the background images and the curve [Refer Fig. 12.].

- **Geometrical properties:** Geometrical properties of background image segments and the curve should be identical and congruent (after several transforms). The generated curve or a curve segment should never be the longest line or the shortest line of the image. LineCAPTCHA should safeguard attacks from edge detection, line detection techniques.

### C. Deployment of LineCAPTCHA

Since the background images are not highly detailed, each image is roughly around 5-6KB. If we take 10 images and if 8 images are randomly selected from the database without replacing any image again, we could generate 1,814,400 unique backgrounds which is an efficient statistic in auto generation. In the terms of data transferring between the app and the server the data array lists will be only a couple of 100 Bytes, where even an average connection speed mobile device will have the chance of using LineCAPTCHA.

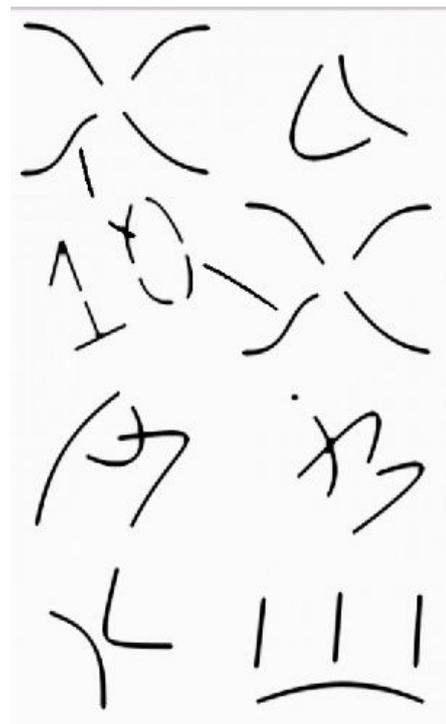
Fig.12. Erosion on short-LineCAPTCHA

V. CONCLUSIONS AND FUTURE WORK

In this paper we have proposed LineCAPTCHA for mobile, an alternative reverse Turing test that will reduce frustrating moments in solving unfriendly CAPTCHAs. It is essential to facilitate CAPTCHA mechanisms in mobile devices since currently mobile devices are overpowering PCs as consumer's primary device of accessing the Internet. The main target of LineCAPTCHA is to improve user-friendliness of reverse Turing tests while maintaining their robustness. We believe users will get more comfortable with LineCAPTCHA as it is easier to draw a line than deciphering unclear text. Domain who can use LineCAPTCHA without getting uncomfortable is comparably higher than most of the other CAPTCHAs since LineCAPTCHA does not depend on age, fluency in English, and even computer literacy. Users can start drawing from anywhere in the curve which is another good attribute in enhancing user-friendliness. LineCAPTCHA proves to be a gateway for humans to pass through within a few of seconds if the users are adapted into LineCAPTCHA. Even though the target was for mobile phone users, LineCAPTCHA can be used efficiently in other devices such as PCs and tablets.

In comparison, this system does not necessitate large database requirements like in Confident CAPTCHA, breakability is much lesser than Confident CAPTCHA (around 3-4%) and compatible with mobile devices in terms of factors such as battery, and display, unlike NuCAPTCHA, Asirra and Vappic.

In future work, when creating images it is even possible to use simple skeleton images of physical objects which are commonly known by most of the users. That will extend the variability of the background images while strengthening the robustness.

Alternative evaluation criteria of LineCAPTCHA can be developed under subject areas such as Artificial Neural Networks and Non-linear regression analysis. We propose obtaining a fitted equation from the curve data set and then predicting y values using that equation through the x values from the draw set then analysing the coincidence of the curve y and predicted y points for verification.

Datasets can be transmitted into a statistical mobile application of statistical software such as Minitab, R or SAS which will lead to verifying both two sample z test and regression analysis methods.